\documentclass[12pt]{article}
\newenvironment{proof}
{\pagebreak[1]{\narrower\noindent {\bf Proof:\quad\nopagebreak}}}{\QED}

\usepackage{amsmath}
\usepackage{amssymb}
\usepackage{url}
\usepackage{graphics}
\usepackage{enumerate}
\usepackage{turnstile}
\usepackage{centernot}

\newcommand{\ang}[1]{\langle#1\rangle}

\newcommand{\xvec}[1]{\ifcase 3{#1} {\ang {x_1,x_2,x_3} } \else
\ifcase 4{#1} {\ang{x_1,x_2,x_3,x_4}} \else {\ang {x_1,\ldots,x_{#1}}}\fi\fi}
\newcommand{\yvec}[1]{\ifcase 3{#1} {\ang {y_1,y_2,y_3} } \else
\ifcase 4{#1} {\ang{y_1,y_2,y_3,y_4}} \else {\ang {y_1,\ldots,y_{#1}}}\fi\fi}
\newcommand{\zvec}[1]{\ifcase 3{#1} {\ang {z_1,z_2,z_3} } \else
\ifcase 4{#1} {\ang{z_1,z_2,z_3,z_4}} \else {\ang {z_1,\ldots,z_{#1}}}\fi\fi}
\newcommand{\vecc}[2]{\ifcase 3{#2} {\ang { {#1}_1,{#1}_2,{#1}_3 } } \else
\ifcase 4{#1} {\ang { {#1}_1,{#1}_2,{#1}_3,{#1}_{4} } }
\else {\ang { {#1}_1,\ldots,{#1}_{#2}}}\fi\fi}
\newcommand{\veccd}[3]{\ifcase 3{#2} {\ang { {#1}_{{#3}1},{#1}_{{#3}2},{#1}_{{#3}3} } } \else
\ifcase 4{#1} {\ang { {#1}_{{#3}1},{#1}_{{#3}2},{#1}_{#3}3},{#1}_{{#3}4} }
\else {\ang { {#1}_{{#3}1},\ldots,{#1}_{{#3}{#2}}}}\fi\fi}
\newcommand{\veccz}[2]{\ifcase 3{#2} {\ang { {#1}_0,{#1}_2,{#1}_3 } } \else
\ifcase 4{#1} {\ang { {#1}_0,{#1}_2,{#1}_3,{#1}_{4} } }
\else {\ang { {#1}_0,\ldots,{#1}_{#2}}}\fi\fi}
\newcommand{\xve}[1]{\ifcase 3{#1} {x_1,x_2,x_3} \else
\ifcase 4{#1} {x_1,x_2,x_3,x_4} \else {x_1,\ldots,x_{#1}}\fi\fi}
\newcommand{\yve}[1]{\ifcase 3{#1} {y_1,y_2,y_3} \else
\ifcase 4{#1} {y_1,y_2,y_3,y_4} \else {y_1,\ldots,y_{#1}}\fi\fi}
\newcommand{\zve}[1]{\ifcase 3{#1} {z_1,z_2,z_3} \else
\ifcase 4{#1} {z_1,z_2,z_3,z_4} \else {z_1,\ldots,z_{#1}}\fi\fi}
\newcommand{\ve}[2]{\ifcase 3#2 {{#1}_1,{#1}_2,{#1}_3} \else
\ifcase 4#2 {{#1}_1,{#1}_2,{#1}_3,{#1}_{4}}
\else {{#1}_1,\ldots,{#1}_{#2}}\fi\fi}
\newcommand{\ved}[3]{\ifcase 3#2 {{#1}_{{#3}1},{#1}_{{#3}2},{#1}_{{#3}3}} \else
\ifcase 4#2 {{#1}_{{#3}1},{#1}_{{#3}2},{#1}_{{#3}3},{#1}_{{#3}4}}
\else {{#1}_{{#3}1},\ldots,{#1}_{{#3}{#2}}}\fi\fi}
\newcommand{\fuve}[3]{
\ifcase 3#2
{{#3}({#1}_1),{#3}({#1}_2,{#3}({#1}_3)} \else
\ifcase 4#2
{{#3}({#1}_1),{#3}({#1}_2),{#3}({#1}_3),{#3}({#1}_4)}
\else
{{#3}({#1}_1),\ldots,{#3}({#1}_{#2})}\fi\fi}
\newcommand{\setmathchar}[1]{\ifmmode#1\else$#1$\fi}
\newcommand{\vlist}[2]{%
	\setmathchar{%
		\compound#2\one{#2}\two
		\ifcompound
			({#1}_1,\ldots,{#1}_{#2})
		\else
			\ifcat N#2
				({#1}_1,\ldots,{#1}_{#2})
			\else
				\ifcase#2
					({#1}_0)\or
					({#1}_1)\or
					({#1}_1,{#1}_2)\or
					({#1}_1,{#1}_2,{#1}_3)\or
					({#1}_1,{#1}_2,{#1}_3,{#1}_4)\else
					({#1}_1,\ldots,{#1}_{#2})
				\fi
			\fi
		\fi}}

\newif\ifcompound
\def\compound#1\one#2\two{%
	\def\one{#1}
	\def\two{#2}
	\if\one\two
		\compoundfalse
	\else
		\compoundtrue
	\fi}

\newcommand{\xwe}[1]{\ifcase 3{#1} {x_1\wedge x_2\wedge x_3} \else
\ifcase 4{#1} {x_1\wedge x_2\wedge x_3\wedge x_4} \else {x_1\wedge\citedots \wedge
x_{#1}}\fi\fi}
\newcommand{\we}[2]{\ifcase 3#2 {\ang { {#1}_1\wedge {#1}_2\wedge {#1}_3 } } \else
\ifcase 4{#1} {\ang { {#1}_1\wedge {#1}_2\wedge {#1}_3\wedge {#1}_{4} } }
\else {\ang { {#1}_1\wedge\citedots\wedge {#1}_{#2}}}\fi\fi}
\newcommand{\s}[1]{\s_{#1}}

\newcommand{\monus}{\;\raise.5ex\hbox{{${\buildrel
    \ldotp\over{\hbox to 6pt{\hrulefill}}}$}}\;}

\newcounter{savenumi}

\newtheorem{theoremfoo}{Theorem}[section] 
\newenvironment{theorem}{\pagebreak[1]\begin{theoremfoo}}{\end{theoremfoo}}

\newtheorem{lemmafoo}[theoremfoo]{Lemma}
\newenvironment{lemma}{\pagebreak[1]\begin{lemmafoo}}{\end{lemmafoo}}
\newtheorem{conjecturefoo}[theoremfoo]{Conjecture}

\newenvironment{conjecture}{\pagebreak[1]\begin{conjecturefoo}}{\end{conjecturefoo}}

\newtheorem{conventionfoo}[theoremfoo]{Convention}

\newtheorem{porismfoo}[theoremfoo]{Porism}

\newtheorem{gamefoo}[theoremfoo]{Game}

\newtheorem{corollaryfoo}[theoremfoo]{Corollary}
\newenvironment{corollary}{\pagebreak[1]\begin{corollaryfoo}}{\end{corollaryfoo}}

\newtheorem{claimfoo}[theoremfoo]{Claim}

\newtheorem{openfoo}[theoremfoo]{Open Problem}

\newtheorem{exercisefoo}{Exercise}

\newcommand{\fig}[1] 
{
 \begin{figure}
 \begin{center}
 \input{#1}
 \end{center}
 \end{figure}
}

\newtheorem{potanafoo}[theoremfoo]{Potential Analogue}

\newtheorem{notefoo}[theoremfoo]{Note}

\newtheorem{notabenefoo}[theoremfoo]{Nota Bene}

\newtheorem{nttn}[theoremfoo]{Notation}

\newtheorem{empttn}[theoremfoo]{Empirical Note}

\newtheorem{examfoo}[theoremfoo]{Example}

\newtheorem{dfntn}[theoremfoo]{Definition}

\newtheorem{propositionfoo}[theoremfoo]{Proposition}

\newcommand{\yyskip}{\penalty-50\vskip 5pt plus 3pt minus 2pt}
\newcommand{\blackslug}{\hbox{\hskip 1pt
        \vrule width 4pt height 8pt depth 1.5pt\hskip 1pt}}
\newcommand{\QED}{{\penalty10000\parindent 0pt\penalty10000
        \hskip 8 pt\nolinebreak\blackslug\hfill\lower 8.5pt\null}
        \par\yyskip\pagebreak[1]}

\newcommand{\BBB}{{\penalty10000\parindent 0pt\penalty10000
        \hskip 8 pt\nolinebreak\hbox{\ }\hfill\lower 8.5pt\null}
        \par\yyskip\pagebreak[1]}

\newtheorem{factfoo}[theoremfoo]{Fact}

\newenvironment{block}{\begin{list}{\hbox{}}{\leftmargin 1em
    \itemindent -1em \topsep 0pt \itemsep 0pt \partopsep 0pt}}{\end{list}}

\usepackage{enumitem}
\usepackage{xcolor}
\usepackage[all]{xy}
\usepackage{wrapfig}

\title{Hardness of Ruling Out Short Proofs of Kolmogorov Randomness}
\author{Hunter Monroe}
\date{\today}
\begin{document}
\maketitle
\begin{abstract}
A meta-complexity assumption, Feasible Chaitin Incompleteness (FCI), asserts the hardness of ruling out length $t$ proofs that string $x$ is Kolmogorov random (e.g. $x{\in}R$), by analogy to Chaitin's result that proving $x{\in}R$ is typically impossible. By assertion, efficiently ruling out short proofs requires, impossibly, ruling out any proof. FCI has strong implications: (i)~randomly chosen $x$ typically yields tautologies hard with high probability for any given proof system, densely witnessing its nonoptimality; (ii)~average-case impossibility of proving $x{\in}R$ implies average-case hardness of proving tautologies and Feige's hypothesis; and (iii)~a natural language is $\textbf{NP}$-intermediate---the sparse complement of ``$x{\in}R$ lacks a length $t$ proof'' (where $R$'s complement is sparse)---and has $\textbf{P/poly}$ circuits despite not being in $\textbf{P}$.

FCI and its variants powerfully assert: (i)~noncomputability facts translate to hardness conjectures; (ii)~numerous open complexity questions have the expected answers (e.g. non-collapse of $\textbf{PH}$), so one overarching conjecture subsumes many questions; and (iii)~an implicit mapping between certain unprovable and hard-to-prove sentences is an isomorphism.

Further research could relate FCI to other open questions and hardness hypotheses; consider whether $R$ frustrates conditional program logic, implying FCI; and consider whether an extended isomorphism maps any true unprovable sentence to hard-to-prove sentences.
\end{abstract}
\section{Introduction}
We conjecture it is typically hard to rule out any length $t$ proof in a theory (e.g. ZFC) that string $x$ is Kolmogorov random ($x{\in}R$) (precise definitions follow, including which theories do the proving).\footnote{This paper was prepared in honor of past and present faculty of Davidson College, including Hansford Epes, L. Richardson King, Benjamin Klein, and Clark Ross. Comments are appreciated from Pavel Pudlák particularly on Conjecture \ref{mainconjecture}. The ideas in this paper and earlier versions have benefited from discussions with the following: Scott Aaronson, Eric Allender, Olaf Beyersdorff, Ilario Bonacino, Cristian Calude, Yuval Filmus, Bill Gasarch, Valentina Harizonov, Pavel Hrubeš, Rahul Ilango, Russell Impagliazzo, Valentine Kabanets, Mehmet Kayaalp, Yanyi Liu, Ian Mertz, Daniel Monroe, Igor Oliveira, Rahul Santhanam, Till Tantau, Neil Thapen, Luca Trevisan, Avi Wigderson, Ryan Williams, Marius Zimand, and other participants in seminars at George Washington University and Davidson College, the Simons Institute 2023 Meta-Complexity Program, the Computational Complexity Conference 2022, the Workshop on Proof Complexity 2022, and the Conference on Complexity with a Human Face 2022. Remaining errors are my own.} This conjecture is analogous to Chaitin's Incompleteness Theorem that proving $x{\in}R$ is typically impossible. ``Typically'' is shorthand for ``except in finite cases''; for strings, this is equivalent to ``for $x$ sufficiently long''. Ruling out length $t$ proofs, which can be done by exhaustive search, is possible but typically hard under the conjecture. An intuition is that a theory typically lacks any proof of $x{\in}R$, and efficiently ruling out length $t$ proofs plausibly requires this inaccessible fact that no proof exists of any length. This conjecture turns out to be far stronger than immediately apparent: it asserts that a fact about noncomputability answers a question in complexity theory, by mapping a typically impossible task, proving $x{\in}R$, to an allegedly hard task. This paper elaborates the conjecture's immediate implications and this possible linkage between noncomputability and hardness.

Conjectures about the hardness of impossible tasks with an added resource bound, such as recognizing TMs halting within $t$ steps or arithmetic sentences having proof length of at most $t$ symbols, date back to at least 1956 (Tseitin) and 1979 (Friedman and later Pudlák) respectively.\footnote{As described Levin\cite{LevinFundamentalsofComputing2020} and Pudlák\cite{PudlakLogicalFoundations}; the work by Tseitin and Friedman is not readily available.} If these tasks are hard, there are as is well known rich implications for the hardness of $\textbf{NP}$-complete languages,  average-case hardness, and the nonexistence of optimal proof systems.

Our conjecture has stronger implications than a similar conjecture by Kraj\'{\i}\v{c}ek and Pudlák\cite{Krajicek} referring to Gödel’s Second Incompleteness Theorem (a consistent theory cannot prove its own consistency). The typical impossibility of proving $x{\in}R$ by Chaitin's Incompleteness Theorem provides a dense, nonconstructive pool of impossible-to-prove sentences.\footnote{For an overview of Kolmogorov complexity, see Li and Vitanyi\cite{LiVitanyiBook}. There is a rapidly growing recent literature on meta-complexity; see Santhanam\cite{SanthanamMetaComplexity}.} Given a theory or proof system, the parameter $x$ can be adjusted to yield allegedly hard-to-prove sentences for it---a second-mover advantage. Furthermore, $R$ is immune---infinite with no infinite computably enumerable (c.e.) subset. By implication, a Turing machine (TM) or theory can distinguish at best a finite subset of $R$. Although finding $x{\in}R$ is typically noncomputable, a randomly chosen $x$ satisfies $x{\in}R$ with high probability (w.h.p.).

Our proposal for identifying hard tautologies based on typical impossibility of proving $x$ Kolmogorov random is not entirely novel. The literature has explored the hardness of random tautologies; tautologies encoding unprovable sentences in a fixed theory; and strings avoiding a pseudo-random generator's range. Our proposal incorporates elements of all three: families of tautologies encoding Kolmogorov randomness are hard; these employ sentences typically unprovable in any given theory;  and it relies on strings $x{\in}R$ avoiding the range of any short program run forever.

Informally, a string is Kolmogorov random (written $x{\in}R$) if it is incompressible by half, with no short description in the form of a program that prints $x$.\footnote{The definition in terms of incompressibility by half is arbitrary, except for Theorem \ref{thmnpintermediate} which requires logarithmic incompressibility.} Chaitin showed it is typically impossible for any theory with a c.e. set of theorems to prove an arithmetic sentence encoding that a string $x$ (represented as a number) lacks a short description. Otherwise, ``the first length $n$ string that provably has no short description'' would itself be a short description of some string, which is a contradiction. This is the intuition behind Chaitin's Incompleteness Theorem. However, the related task of ruling out any proof of $x{\in}R$ with at most $t$ symbols is possible, by examining each proof, but hard under our conjecture. The conjecture informally asserts that if proving $x{\in}R$ in a theory is impossible, then the theory lacks sufficient information to rule out all length $t$ proofs efficiently---the fact that no proofs exist for any $t$ is inaccessible, by Chaitin's Theorem. Ignorance about $x{\in}R$ would be absolute, with no ``free lunch'' gained from adding the resource bound $t$ that would somehow allow efficient proofs. 

The conjecture has strong implications:
\begin{itemize}\setlength\itemsep{-0.4em}
\item Chaitin's Theorem implies that unprovable sentences have positive density, because a random string satisfies $x{\in}R$ with high probability, and sentence $x{\in}R$ is typically unprovable. If the conjecture holds, the task of ruling out length $t$ proofs is hard on average. Equivalently, proving tautologies encoding this statement is hard, and Feige's hypothesis holds.

\item Chaitin's Theorem implies that no axiom system is optimal, because new axioms $x{\in}R$ can be added---and these are dense. Likewise, under the conjecture, there is no optimal proof system for tautologies, with dense hard families witnessing the nonoptimality.

\item A natural language is $\textbf{NP}$-intermediate: the sparse complement of the language ``$x{\in}R$ lacks a length $t$ proof'' (where $R$ is redefined, by requiring logarithmic incompressibility, to have a sparse complement). This language is not in $\textbf{P}$ but has $\textbf{P/poly}$ circuits.

\item If the conjecture holds for harder versions of $R$ noncomputable even by a TM with an oracle for halting (and applying oracles iteratively), then $\textbf{PH}$ does not collapse. 
\end{itemize}

The conjecture deserves careful study for several reasons. First, it yields new candidates for hard tautologies, potentially hard for any proof system. Second, it concisely asserts that widely-held beliefs about multiple open questions are correct. Finally, it implies there is an isomorphism between certain unprovable and hard-to-prove sentences in a theory such as ZFC, which if extended would be an important structural feature. 

To present the conjecture precisely, define the set of Kolmogorov random strings as $R{=}\{x|\forall p{:}$ if $|p|{\leq}|x|/2$, then $p{\nearrow}$ or $p{\downarrow}$ with $U(p){\neq}x\}$, with $U$ a deterministic universal TM with no limit on its running time (not necessarily prefix free), $x$ and $p$ binary strings with $|x|$ denoting $x$'s length, $p{\downarrow}$ and $p{\nearrow}$ signifying program $p$ does or does not halt, and `$x{\in}R$' is an arithmetic sentence encoding $x{\in}R$. Single and double quotes signify a sentence, a sequence of symbols, encoding a mathematical statement. Proof length is the length in symbols of the string representing the proof.\footnote{See Pudlák\cite{PudlakLengthsOfProofs}'s survey on proof length.}

Formalizing the intuition ``ruling out length $t$ proofs is hard'' requires specifying which theory lacks length $t$ proofs and which theory has difficulty verifying this. These theories must be different theories for the following reason. An inconsistent theory has short indirect proofs of any sentence, even false ones: $\neg\phi$ implies a contradiction, so $\phi$ is proven. Therefore, if a theory proves it lacks short proofs of any $\phi$, it proves its own consistency. By analogy with Chaitin's theorem, we conjecture: 

\begin{conjecture}\label{mainconjecture}
For every consistent theory $\mathcal{S}$, there exists a consistent theory $\mathcal{T}$, such that for all $x{\in}R$:
$\mathcal{S}$ typically lacks $t^{\mathcal{O}(1)}$ length proofs of ``$\mathcal{T}$ lacks any proof of `$x{\in}R$' within length $t${''} (e.g. $\mathcal{S}\,{\centernot{\sststile{}{t^{\mathcal{O}(1)}}}}``\mathcal{T}\,{\centernot{\sststile{}{t}}}[`x{\in}R$'$\,]${''}). 
\end{conjecture}

In the notation above in parentheses, write $\mathcal{T}\,{\sststile{}{}}\phi$ or $\mathcal{T}\,{\centernot{\sststile{}{}}}\phi$ respectively if $\mathcal{T}$ does or does not have a proof of $\phi$ of any length. Write $\mathcal{T}\,{\sststile{}{t}}\phi$ if theory $\mathcal{T}$ has a length $t$ (or shorter) proof of sentence $\phi$ and $\mathcal{T}\,{\centernot{\sststile{}{t}}}\phi$ if not. Likewise, $\mathcal{T}\,{\centernot{\sststile{}{t^{\mathcal{O}(1)}}}}\phi$ signifies that $\mathcal{T}$ does not have an efficient (polynomially bounded) proof of $\phi$. $\mathcal{T}\,{\sststile{}{}}\phi$ and $\mathcal{T}\,{{\sststile{}{\mathcal{O}(1)}}}\phi$ are equivalent; a provable sentence has a finite proof and is therefore provable within a constant bound. The square brackets in ``$\mathcal{T}{\centernot{\sststile{}{t}}}[`x{\in}R$'$\,]${''} are included only for readability.

Conjecture \ref{mainconjecture} has similar implications for the length of proofs of tautologies (under the encoding in Section \ref{sectiontautologies}) in proof systems:

\begin{corollary}\label{corollarytautology}
Under Conjecture \ref{mainconjecture}, tautologies encoding ``$\mathcal{T}\,{\centernot{\sststile{}{t}}}[`x{\in}R$'$\,]${''} typically lack $t^{\mathcal{O}(1)}$ length proofs in proof system $P$. 
\end{corollary}
\begin{proof}
Choose a theory $\mathcal{S}$ that proves the sentence `$P$ is sound'. Choose $\mathcal{T}$ and $x$ by Conjecture \ref{mainconjecture}. Then tautologies encoding ``$\mathcal{T}{\centernot{\sststile{}{t}}}[`x{\in}R$'$\,]${''} must typically lack $t^{\mathcal{O}(1)}$ length proofs in any proof system $P$. Otherwise, $\mathcal{S}$ could ``import'' $P$'s efficient proof if one existed. More precisely, $\mathcal{S}$ would have a $t^{\mathcal{O}(1)}$ length proof that consists of $P$'s $t^{\mathcal{O}(1)}$ length proof combined with the proof in $\mathcal{S}$ that `$P$ is sound'. 
\end{proof}	

Conjecture \ref{mainconjecture} builds on the literature  exploring relationships between what is impossible to prove and hard to prove in a given theory. Pudlák states a Feasible Incompleteness Thesis: ``The phenomenon of incompleteness manifests itself at the level of polynomial time computations''.\footnote{See Pudlák\cite{PudlakLogicalFoundations} Section 6.4 and \cite{PudlakFiniteDomain}.} One particular impossible task, by Gödel’s Second Incompleteness Theorem, is for a consistent theory to prove its own consistency. Pudlák and Friedman independently formulated a ``feasible consistency'' conjecture: a consistent theory $\mathcal{T}$ cannot efficiently show there is no proof of contradiction derivable from the axioms of $\mathcal{T}$ within length $t$, e.g., $\mathcal{T}\centernot{{{\sststile{}{t^{\mathcal{O}(1)}}}}}$``$\mathcal{T}\,{\centernot{\sststile{}{t}}}[`0{=}1$'\,]{''}. Pudlák\cite{Pudlak1986length} shows this initial conjecture is incorrect.\footnote{See also Theorem 59 of Pudlák\cite{PudlakLogicalFoundations}.} A reformulated conjecture, by Kraj\'{\i}\v{c}ek and Pudlák\cite{Krajicek}, is that for any consistent theory $\mathcal{S}$, there exists a (stronger) consistent theory $\mathcal{T}$ such that  $\mathcal{S}\,{{\centernot{\sststile{}{t^{\mathcal{O}(1)}}}}}$``$\mathcal{T}\,{\centernot{\sststile{}{t}}}[`0{=}1$'\,]{''}. They prove this is equivalent to the conjecture that no optimal propositional proof system exists, which remains a key open question in proof complexity. Our conjecture differs by replacing the sentence `$0{=}1$' with a sentence `$x{\in}R$', where $x$ is chosen to be sufficiently long that the sentence is unprovable in both $\mathcal{S}$ and $\mathcal{T}$. Our stronger conjecture, referring to a dense nonconstructive set of unprovable sentences, not surprisingly has stronger implications than the nonexistence of optimal proof systems, as described above.

Conjecture \ref{mainconjecture} refers to meta-complexity---the complexity of complexity---in multiple ways.\footnote{For an introduction to meta-complexity, see Santhanam\cite{SanthanamMetaComplexity}.} It employs complexity of proving Kolmogorov complexity. It employs incompleteness theorems limiting what a theory can prove especially about itself. It also turns proof complexity in one theory against proof complexity in another, and links impossibility and hardness. This paper also addresses several concerns in the meta-complexity literature: reducing average-case hardness to a weaker assumption such as worst-case hardness (or in our case, a fact about average-case impossibility), determining whether certain languages are $\textbf{NP}$-hard or $\textbf{NP}$-intermediate, and consolidating multiple meta-complexity hypotheses into a single unifying hypothesis. This paper does not address how our conjecture relates to the meta-complexity literature's hypotheses.\footnote{These concern for instance the average-case hardness of time-bounded Kolmogorov complexity (Liu and Pass\cite{LiuPass}) and the hardness of distinguishing Kolmogorov random strings from other strings under samplable distributions (Ilango et al\cite{Ilangoetall}). Our resource bound is on proof length, while theirs is on the time allowed for a short program $p$ to output $x$ and the resources allowed to sample a distribution respectively.}

The paper is organized as follows. Section \ref{preliminariessection} provides preliminaries. Section \ref{sectioncaludejurgensen} shows that unprovable sentences `$x{\in}R$' are dense among length $n$ sentences. Section \ref{sectiontautologies} discusses implications proof systems. Section \ref{sectionmapping} shows that the mapping from impossible to hard tasks is an isomorphism and reinterprets previous results in that light. Section \ref{sectionexploring} presents stronger conjectures that would extend the isomorphism. Section \ref{conclusion} concludes.

\section{Preliminaries}\label{preliminariessection}
\emph{Strings}: With a binary alphabet $\{0,1\}$, let $S^n$ be the set of length $n$ strings, which are ordered $n$-tuples. Let $|x|$ be the length of a string and $|S|$ be the cardinality of set $S$. A language $L$ is a subset of $\cup_{n\geq 0}S^n$.

\emph{Density}: Say the share of length $n$ strings in $L$ is bounded above zero if there exists $c>0$ such that $|L\cap S^n|/n\geq c$  for sufficiently large $n$. This implies the weaker condition that $L$ has positive upper density, i.e., that $\displaystyle\limsup_{n \rightarrow\infty}\frac{|L\cap\{1,2,\ldots,n\}|}{n}>0$. If an event depending on $n$ occurs with probability that tends to one as $n$ tends to infinity, such as $x{\in}R$ where $|x|{=}n$, say that it occurs with high probability (w.h.p.).

\emph{Theories}: Theories are assumed to be the Peano arithmetic (PA) or an extension of PA.\footnote{The conjecture could coherently refer to a weaker theory such as Robinson's Q without induction or unbounded quantifiers, which can still prove `$p{\downarrow}$' if in fact $p{\downarrow}$, by verifying the transcript of a halting computation.} To allow for average-case analysis, the standard definition of PA is modified so binary strings are encoded in arithmetic sentences as natural numbers, in binary not unary, adding a leading ``1'' to avoid losing leading zeros.

\emph{Proof Systems}: A propositional proof system is a polynomial time function $h\in \textbf{FP}$ with range $\texttt{TAUT}$ (Cook and Reckhow\cite{CookReckhow}). For tautology $\tau$, any string $w$ such that $h(w)=\tau$ is a proof of $\tau$. The proof system $h$ is \emph{optimal} if there exists $c\geq 1$ such that the length of minimal $f$ proofs of $x$ are polynomially bounded in $|x|$ with exponent $c$ by minimal $h$ proofs (Kraj\'{\i}\v{c}ek and Pudlák\cite{Krajicek}).

\section{Density of Unprovable Sentences}\label{sectioncaludejurgensen}
Calude and J{\"u}rgensen\cite{CaludeJurgensen} show that the share of length $n$ arithmetic sentences that are true and unprovable is bounded above zero. The result relies on two facts: `$x{\in}R$' is typically unprovable, and length $n$ strings are in $R$ w.h.p.\footnote{See the proof of Theorem 5.2 in \cite{CaludeJurgensen}.} With that context, Conjecture \ref{mainconjecture} states that a similar result holds for $\texttt{coTHEOREMS}_{\leq t}{=}$ $\{\langle \phi,1^t\rangle|\mathcal{T}\,{\centernot{\sststile{}{t}}}\phi\}$.

Chaitin’s Incompleteness Theorem states:
\begin{theorem}\label{chaitinthm}
	For every consistent, arithmetically sound theory $\mathcal{T}$ with a c.e. set of theorems, $\exists k\forall x{:}|x|{>}k$, $\mathcal{T}\,{\centernot{\sststile{}{}}}`x{\in}R$'.
\end{theorem}
\begin{proof}
Otherwise, a string $x$ could be concisely described as ``the first string $x$ of length $n$ such that $\mathcal{T}$ proves `$x{\in}R$''', contrary to the definition of $R$. A TM with input $n$ in binary (of length $\log n$) could enumerate the theorems of $\mathcal{T}$, printing the first string $x$ such that $\mathcal{T}$ proves `$x{\in}R$'. Then, $k$ is determined by the length of the description of this TM, which would need to be doubled as $R$ consists of strings not compressible by half. See Li and Vitanyi\cite{LiVitanyiBook} Corollary 2.7.2 for a formal treatment.
\end{proof}

\begin{lemma}\label{lemmahighprobabilityR}
$x{\in}R$ w.h.p.
\end{lemma}
\begin{proof}
By a counting argument, the number of possible short descriptions is small. The number of length $n$ strings is $2^n$. The number of programs $p$ with $|p|{\leq}n/2$ is $2^{n/2+1}-1$, which is an upper bound on the number of length $n$ strings not in $R$. Therefore, $R$'s share of length $n$ strings is at least $1-2^{-n/2}$, so $x{\in}R$ w.h.p.
\end{proof}

Calude and J{\"u}rgensen's result implies:
\begin{theorem}\label{CaludeJurgensenInformal}
For every theory $\mathcal{T}$, the share of sentences $\{`x{\in}R$'$\,|$ $x{\in}R$ and $\mathcal{T}\,{\centernot{\sststile{}{}}}`x{\in}R$'$\}$ in length $n$ arithmetic sentences is bounded above zero, for $n$ sufficiently large. 
\end{theorem}
\begin{proof}
Theory $\mathcal{T}$ cannot typically prove sentences `$x{\in}R$' where $x{\in}R$, by Theorem \ref{chaitinthm}.
The sentences `$x{\in}R$' satisfy $|`x{\in}R$'$|=|x|+c$, where $c$ is a constant not depending on $|x|$, giving the overhead of encoding `$x{\in}R$' net of $|x|$. The share of length $n$ sentences of form `$x{\in}R$' is exactly $2^{-c}$ and these satisfy $x{\in}R$ w.h.p. Therefore, for $\epsilon{>}0$, this share is bounded below by $2^{-c}{-}\epsilon$ for $n$ sufficiently large.
\end{proof}
The fact that a sentence `$x{\in}R$' needs only a constant $c$ bits of overhead, net of $|x|$, to encode $x{\in}R$ is needed in the next section.
\section{Tautologies and Proof Systems}\label{sectiontautologies}

A tautology can encode the sentence ``$\mathcal{T}{\centernot{\sststile{}{t}}}[`x{\in}R$'$\,]${''} as follows. For a given $x$, $\mathcal{T}{\centernot{\sststile{}{t}}}`x{\in}R$' is equivalent to $\langle `x{\in}R$'$,1^t\rangle{\in}\texttt{coTHEOREMS}_{\leq t}$. $\texttt{coTHEOREMS}_{\leq t}$ and $\texttt{TAUT}$ are both  $\textbf{coNP}$-complete languages, so some polynomial-time reduction $r$ from $\texttt{coTHEOREMS}_{\leq t}$ to $\texttt{TAUT}$ maps $\langle \phi,1^t\rangle$ to tautology $r(\langle \phi,1^t\rangle)$. Conjecture \ref{mainconjecture} implies that for any proof system $P$, for $x{\in}R$, the family of tautologies $r(\langle `x{\in}R$'$,1^t\rangle)$ typically lacks $t^{\mathcal{O}(1)}$ length proofs. Equivalently, $r(\langle `x{\in}R$'$,1^t\rangle)$ has $t^{\mathcal{O}(1)}$ length proofs in $P$ for only a finite set of $x{\in}R$. 

Tautologies produced by the reduction $r$ confirm that every possible proof of ``$\mathcal{T}{\centernot{\sststile{}{t}}}[`x{\in}R$'$\,]${''} is not a valid proof. The reduction $r$ translates a family of sentences stating that no length $t$ proof exists to a family of tautologies. It should not be confused with propositional translations, which translate sentences with a single universal bounded quantifier that are easy to prove in a weak fragment of arithmetic into easy-to-prove tautologies.\footnote{See Kraj\'{\i}\v{c}ek\cite{Krajicekproof} and Cook and Nguyen\cite{CookNguyen}.} For instance, a propositional translation may associate each bit in a $k$-bit number with a logical variable in a Boolean formula, which is a tautology if the sentence is true for all $k$-bit numbers. $r$ can be seen as a dual approach to propositional translation differing in every respect. Its domain is unprovable, not provable sentences. Its range is allegedly hard tautologies, rather than provably easy tautologies. It accommodates unbounded quantifiers with alternation, rather than just a single bounded universal quantifier. Its range is tautologies encoding ``no length $t$ proof in $\mathcal{T}$ is valid'', rather than ``$\phi$ holds for all $k$-bit numbers''. The reduction $r$ and propositional translation can be seen as dual, as they differ in every respect.

With this encoding, three implications immediately follow:  $\texttt{TAUT}{\notin}\textbf{AvgP}$; Feige's hypothesis; and there are dense witnesses to the nonoptimality of proof systems. 

\subsection{$\texttt{TAUT}{\notin}\textbf{AvgP}$}
$R$'s density implies $\texttt{TAUT}{\notin}\textbf{AvgP}$ by Conjecture \ref{mainconjecture} and Theorem \ref{CaludeJurgensenInformal}. Recall that a distributional problem $(L,\mathcal{D})$ (where $\mathcal{D}$ is an ensemble of distributions for length $n$ inputs) is in $\textbf{AvgP}$ if there is a TM $M$ accepting $L$ and constants $C$ and some $\epsilon>0$ such for that every $n$: $\displaystyle \mathop{\mathbb{E}}_{y\in_R \mathcal{D}_n} \displaystyle\bigg[ \frac{T_M(y)^\epsilon}{n} \bigg]\leq C$, where $T_M$ is the function that maps a string $x$ to how many steps $M(x)$ takes, and $\mathbb{E}$ is the expectation operator.

\begin{theorem}\label{theoremavgp}
Under Conjecture \ref{mainconjecture}: (i) $\texttt{coTHEOREMS}_{\leq t}{\notin}\textbf{AvgP}$ and (ii) $\texttt{TAUT}{\notin}\textbf{AvgP}$, for any distribution applying non-negligible weight to elements of the form $\langle `x{\in}R$'$,1^t\rangle$ and $r(\langle `x{\in}R$'$,1^t\rangle)$.
\end{theorem}
\begin{proof}
(i) Conjecture \ref{mainconjecture} implies that the share of length $n$ elements in the language $\texttt{coTHEOREMS}_{\leq t}$ of the form $\langle `x{\in}R$'$,1^t\rangle$ is bounded above zero, which is sufficient to show $\texttt{coTHEOREMS}_{\leq t}{\notin}\textbf{AvgP}$, by the following argument similar to the proof of Theorem \ref{CaludeJurgensenInformal}. That proof notes that $|$`$x{\in}R$'$|{=}|x|{+}c$. Therefore, $|\langle `x{\in}R$'$,1^t\rangle|{=}|x|{+}c{+}t$ for a slightly larger $c$. Fix $n$, let $|x|$ range from 0 to $n{-}c$, and let $t=n{-}|x|{-}c$. All but $x$ is fixed, so among length $n$ elements, the number of possibilities for $x$ in inputs $\langle `x{\in}R$'$,1^t\rangle$ of length $n$ is  $2^0{+}\ldots{+}2^{n-c}{=}2^{n-c+1}{-}1$. Any string $x$ is in $R$ w.h.p. by Lemma \ref{lemmahighprobabilityR}. Similarly, the share of length $n$ elements of the form $\langle `x{\in}R$'$,1^t\rangle$ with $x{\in}R$ is roughly $2^{n-c+1}/2^n$, so the share exceeds $2^{-c+1}{-}\epsilon$ for some $\epsilon{>}0$ for $n$ sufficiently large. 

(ii) The same argument holds for tautologies $r(\langle `x{\in}R$'$,1^t\rangle)$, if $r$ is chosen to map all elements in its domain of length $n$ to tautologies all of the same length, say, $64n$, implying the share of length $n$ tautologies of the form $r(\langle `x{\in}R$'$,1^t\rangle)$ with $x{\in}R$ is bounded above zero by $2^{-c+1-64}{-}\epsilon$ for some $\epsilon{>}0$ for $n$ sufficiently large. 
\end{proof}

\subsection{Feige's Hypothesis}
In a random $\texttt{k-CNF}$, with $n$ Boolean variables and $k$ variables per clause, there are ${n\choose k}$ choices for variables in a clause, each of which may or may not be negated. Suppose a random $\texttt{k-CNF}$ consists of a fraction $\Delta n$ from these possibilities for a constant $\Delta>0$. For $\Delta$ above an established threshold, the formula is unsatisfiable w.h.p. Feige\cite{Feige02} conjectures that $\texttt{k-CNF}$ refutation is hard above this threshold. The set $\texttt{UNSAT}_k$ of unsatisfiable $\texttt{k-CNF}$s is $\textbf{coNP}$-complete, so there is a reduction $u$ to this language from $\texttt{coTHEOREMS}_{\leq t}$. Again choose $u$ to map all elements in its domain of length $n$ to elements of the same length in $\texttt{UNSAT}_k$.

\begin{theorem}\label{theoremfeige}
Conjecture \ref{mainconjecture} implies that Feige's hypothesis holds for a sufficiently high clause density $\Delta$ determined by $u$.
\end{theorem}
\begin{proof}
The share of length $n$ inputs $\langle `x{\in}R$'$,1^t\rangle$ in $\texttt{coTHEOREMS}_{\leq t}$ is bounded above zero by Theorem \ref{theoremavgp}(i). Therefore, the share of length $n$ inputs $u(\langle `x{\in}R$'$,1^t\rangle){\in}\texttt{UNSAT}_k$ is also bounded above zero. The reduction $u$ will imply that these $\texttt{k-CNF}$s have a certain clause density $\Delta$. Feige's Hypothesis holds at this clause density. By changing the choice of $u$ to pad the length of its output, Feige's hypothesis also holds at a higher fixed clause density (with a lower share of hard tautologies).
\end{proof}
\subsection{Dense Witnesses to Nonoptimality}
Conjecture \ref{mainconjecture} implies there is a dense set of hard families of tautologies $r(\langle `x{\in}R$'$,1^t\rangle)$ letting $x$ range over all $x{\in}R$. A probabilistic, polynomial-time computable procedure to produce such a family w.h.p. is to choose a sufficiently long random string $x$. Then, $x{\in}R$ w.h.p. by Lemma \ref{lemmahighprobabilityR}, so tautologies $r(\langle `x{\in}R$'$,1^t\rangle)$ are hard for $P$ w.h.p. Tautologies that are hard for ZFC to prove are also hard for any other known proof system, as their soundness is proved by ZFC. Conjecture \ref{mainconjecture} does not tie down how long is ``sufficiently long''. Conjecture \ref{conjectureisomorphism} implies that ``sufficiently long'' is the same as $k$ in Chaitin's theorem. This is based on the length of the description of a TM that enumerates the theorems of a theory. 

\section{Mapping Impossible to Hard Tasks}\label{sectionmapping}
This section shows that the mapping between impossible-to-prove sentences (`$x{\in}R$') and hard-to-prove sentences (no length $t$ proof of `$x{\in}R$' exists) implicit in Conjecture \ref{mainconjecture} is in fact an isomorphism. It reinterprets each of the above results as translating facts about noncomputability directly into hardness conjectures using this isomorphism. It then proposes some indirect tests of this interpretation, which are addressed in the next section.
\subsection{Impossible and Hard Tasks are Isomorphic}
To spell out Conjecture \ref{mainconjecture}'s implied translation of impossible into hard tasks, define the mapping $\mathfrak{R}$ for all $x$ (not only those satisfying $x{\in}R$), from the sentence `$x{\in}R$' to the family of sentences ``$\mathcal{T}\,{\centernot{\sststile{}{t}}}[`x{\in}R$'$\,]${''} indexed by $t$. Sentences in the domain of $\mathfrak{R}$ may be impossible or possible to prove/refute in $\mathcal{S}$, while families in the range may be hard or easy to prove/refute in $\mathcal{S}$. Conjecture \ref{mainconjecture} has the implies that $\mathfrak{R}$ is an isomorphism: 

\begin{theorem}\label{theoremisomorphism} 
	$\mathfrak{R}$ maps impossible to hard tasks, and possible to easy tasks, under Conjecture \ref{mainconjecture}.\footnote{A possible task $\mathcal{T}\,{\sststile{}{}}\phi$ and an easy task $\mathcal{T}\,{{\sststile{}{\mathcal{O}(1)}}}\phi$ are the same thing as noted above.}
\end{theorem}
\begin{proof}
	Consider three cases: (i) if  $\mathcal{S}\,{\centernot{\sststile{}{}}}`x{\in}R$' and $x{\in}R$, then by Conjecture \ref{mainconjecture}, $\mathcal{S}\,{\centernot{\sststile{}{t^{\mathcal{O}(1)}}}}\mathfrak{R}(`x{\in}R$'$)$. Therefore, $\mathfrak{R}$ maps an impossible task (proving `$x{\in}R$') to a hard one (proving $\mathfrak{R}(`x{\in}R$'$)$).
	
	(ii) If $\mathcal{S}\,{{\sststile{}{}}}`x{\in}R$', $\mathfrak{R}$ maps a possible task (proving `$x{\in}R$') to a easy one, since refuting ``$\mathcal{T}\,{\centernot{\sststile{}{t}}}[`x{\in}R$'$\,]${''} is easy for $t$ sufficiently large for $\mathcal{S}$ to verify the proof of $`x{\in}R$'.
	
	(iii) If $x{\notin}R$, then $\mathcal{S}\,{{\sststile{}{}}}`x{\notin}R$' and therefore $\mathcal{S}\,{{\sststile{}{\mathcal{O}(1)}}}`x{\notin}R$' because the counterexample $p$ with $U(p){=}x$ is easy to verify---see Lemma \ref{lemmahalting}. Then $\mathfrak{R}$ maps a possible task (refuting `$x{\in}R$') to an easy task (refuting `$x{\in}R$' within a constant bound).
\end{proof}

\begin{lemma}\label{lemmahalting}
	For any consistent theory $\mathcal{T}$ that can encode computation, if $p{\downarrow}$, $p{\downarrow}$ with $U(p){\neq}x$, or $x{\notin}R$, then $\mathcal{T}\,{{\sststile{}{}}}`p{\downarrow}$', $\mathcal{T}\,{{\sststile{}{}}}`U(p){\neq}x$', and $\mathcal{T}\,{{\sststile{}{}}}`x{\notin}R$' respectively, without use of induction or unbounded quantifiers, and therefore $\mathcal{T}\,{{\sststile{}{\mathcal{O}(1)}}}`p{\downarrow}$', $\mathcal{T}\,{{\sststile{}{\mathcal{O}(1)}}}`U(p){\neq}x$', and $\mathcal{T}\,{{\sststile{}{\mathcal{O}(1)}}}`x{\notin}R$'. 
\end{lemma}
\begin{proof}
	If $p{\downarrow}$, then $\mathcal{T}\,{{\sststile{}{}}}`p{\downarrow}$', with the computation's transcript as proof. Induction and unbounded quantifiers are not required to prove that a program halts (Papadimitriou\cite{Papadimitriou} Lemma 6.1). The length of the proof is bounded above by a constant, so $\mathcal{T}\,{{\sststile{}{\mathcal{O}(1)}}}`p{\downarrow}$'. The same argument applies for the remaining two cases.\footnote{There is a dichotomy: `$x{\in}R$' is hard to prove or is refutable with a constant upper bound.}
\end{proof}

\subsection{Translating Noncomputability Facts into Hardness Conjectures}
This section argues that $\mathfrak{R}$ not only maps impossible to hard tasks, but also translates noncomputability facts into hardness conjectures. The following theorem restates the above implications of Conjecture \ref{mainconjecture} as noncomputability results translated by isomorphism into complexity theory: 

\begin{theorem}\label{proofbyisomorphism}
Theorem \ref{chaitinthm} (Chaitin) implies that there is no optimal theory $\mathcal{T}$, as a new axiom `$x{\in}R$' can always be added. This implies there is no optimal proof system $P$, by the isomorphism under Conjecture \ref{mainconjecture}.

Theorem \ref{CaludeJurgensenInformal} (Calude and J{\"u}rgensen) shows that the share of length $n$ sentences that are true and unprovable is bounded above zero. This implies Theorems \ref{theoremavgp} ($\texttt{TAUT}{\notin}\textbf{AvgP}$) and \ref{theoremfeige} (Feige's Hypothesis), by the isomorphism under Conjecture \ref{mainconjecture}.

Theorem \ref{chaitinthm} provides a procedure to find $x{\in}R$ such that $\mathcal{T}\,{\centernot{\sststile{}{}}}`x{\in}R$' w.h.p. Any $x$ typically yields tautologies hard w.h.p. for proof system $P$ w.h.p. if $\mathcal{T}$ proves that $P$ is sound, by the isomorphism under Conjecture \ref{mainconjecture}.
\end{theorem}

The idea that an isomorphism under Conjecture \ref{mainconjecture} translates computability facts to hardness conjectures is an appealing approach to complexity theory's open questions. It is also appealing that an isomorphism might define an overlooked symmetry between the unprovable and hard-to-prove sentences in a theory such as ZFC. Verifying this idea directly would require proving the conjecture, thereby resolving multiple open problems.

An indirect test is to show that we can extend the isomorphism's reach. This might be done in four ways: (i)~identifying an additional computability fact that is translated into a hardness conjecture; (ii)~identifying an additional hardness conjecture implied by a computability fact; (iii)~identifying larger classes of unprovable sentences beyond $x{\in}R$ that could be translated; and (iv)~identifying overarching conjectures subsuming the examples and suggesting underlying principles. The next section pursues these avenues.

\section{Extending the Isomorphism}\label{sectionexploring}
This section provides indirect evidence of the isomorphism's validity by extending it in the four ways just noted. Section \ref{sectionintermediate} uses the fact that $R$ is Turing intermediate to suggest a language that is $\textbf{NP}$-intermediate. Section \ref{sectionphnoncollapse} extends the isomorphism to imply that neither the arithmetic hierarchy ($\textbf{AH}$) nor the polynomial hierarchy ($\textbf{PH}$) collapse. Section \ref{sectionfoldingin} folds some other open questions into this framework. Section \ref{sectionoverarching} proposes an overarching conjecture subsuming the previous complexity hypotheses. Finally, Section \ref{sectionextendingisomorphism} considers maximal extensions of the isomorphism.
\subsection{From Turing Intermediate to $\textbf{NP}$ Intermediate}\label{sectionintermediate}
The apparent pattern in Theorem \ref{proofbyisomorphism} is that $\mathfrak{R}$ translates facts about noncomputability implied by Chaitin's Theorem into hardness conjectures, so we consider another implication of that Theorem. The set $R$ is Turing intermediate---it is not computable, and its complement is c.e. but not complete under many-one computable reductions (Rogers\cite{Rogers} Theorem 8.I(a) and (c)). This raises the question whether Conjecture \ref{mainconjecture} implies that some related language is $\textbf{NP}$-intermediate---that is, in $\textbf{NP}$, not in $\textbf{P}$, and not $\textbf{NP}$-complete under polynomial time many-one reductions. This section shows this is correct for a variant of Conjecture \ref{mainconjecture}. The final paragraph provides context on $\textbf{NP}$-intermediate languages.

We show that deciding the language ``has no proof of `$x{\in}R$' within length $t$'' is $\textbf{NP}$-intermediate under a stronger version of Conjecture \ref{mainconjecture} that relaxes $R$'s definition to make its complement sparse. This relaxed definition counts strings as random unless they can be compressed exponentially, not just by half. This makes the set of possible short descriptions sparse, growing polynomially in $|x|$, so the the set of non-random strings is also sparse. Define this sparse version of $R$ as $R^{sp}{=}\{x|\forall p{:}$ if $|p|{\leq}\log|x|$, then $p{\nearrow}$ or $p{\downarrow}$ with $U(p){\neq}x\}$. $R^{sp}$, like $R$, is noncomputable. Chaitin's Theorem still holds, but the parameter $k$ is exponentially larger. Suppose Conjecture \ref{mainconjecture} holds using $R^{sp}$ rather than $R$, and fix $\mathcal{S}$ and $\mathcal{T}$. As above, $\mathcal{T}{\centernot{\sststile{}{t}}}`x{\in}R^{sp}$' iff $\langle `x{\in}R^{sp}$'$,1^t\rangle{\in}\texttt{coTHEOREMS}_{\leq t}$, by definition. Let $R^{sp}_t{=}\{\langle `x{\in}R^{sp}$'$,1^t\rangle|$$\mathcal{T}{\centernot{\sststile{}{t}}}`x{\in}R^{sp}$'$\}$, so $R^{sp}_t{\in}\texttt{coTHEOREMS}_{\leq t}$. Define $\overline{R^{sp}_t}{=}\{\langle `x{\in}R^{sp}$'$,1^t\rangle|\neg$$\mathcal{T}{\centernot{\sststile{}{t}}}`x{\in}R^{sp}$'$\}$. Based on $x$, $\overline{R^{sp}_t}$ can be divided into $x{\notin}R^{sp}$ where $\langle `x{\in}R^{sp}$'$,1^t\rangle{\in}\overline{R^{sp}_t}$ for all $t$, and $x{\in}R^{sp}$ where $\langle `x{\in}R^{sp}$'$,1^t\rangle{\in}\overline{R^{sp}_t}$ for sufficiently large $t$. Given that $\overline{R^{sp}_t}$ is sparse, this conjecture will achieve our objective:

\begin{theorem}\label{thmnpintermediate}
If Conjecture \ref{mainconjecture} holds with $R$ replaced with $R^{sp}$: (i) $\overline{R^{sp}_t}$ is $\textbf{NP}$-intermediate; and (ii) $\overline{R^{sp}_t}$ and therefore $R^{sp}_t$ have minimal circuits in $\textbf{P/poly}$ which are not $\textbf{P}$-uniform.
\end{theorem}
\begin{proof}
(i) $R^{sp}{\notin}\textbf{P}$ by assumption. $\overline{R^{sp}_t}$ is sparse, as $R^{sp}$ was defined to ensure this. A sparse language is not $\textbf{NP}$-complete under many-one reductions unless $\textbf{P}{=}\textbf{NP}$, which the assumption rules out (Mahaney\cite{Mahaney82}).

(ii) $\overline{R^{sp}_t}$ is sparse, so it has minimal circuits in $\textbf{P/poly}$. These are not $\textbf{P}$-uniform, which would imply $R^{sp}{\in}\textbf{P}$, which does not hold by assumption.
\end{proof}

Allender and Hirahara\cite{AllenderHiraharaNPIntermediate} also provide examples of natural languages that are conditionally $\textbf{NP}$-intermediate. They show that if one-way functions exist, then approximating minimum circuit size and time-bounded Kolmogorov complexity are $\textbf{NP}$-intermediate. Determining whether the minimum-circuit size problem and related problems on time-bounded Kolmogorov complexity are $\textbf{NP}$-hard or not is an area of active research; see for instance Hirahara\cite{HiraharaMCSPNPComplete}.

Previous results constructed artificial  $\textbf{NP}$-intermediate languages. Ladner\cite{Ladner} shows that if $\textbf{P}{\neq}\textbf{NP}$, there exists an  $\textbf{NP}$-intermediate language, constructed artificially. Mahaney showed that a sparse language is not $\textbf{NP}$-complete under many-one reductions unless $\textbf{P}{=}\textbf{NP}$, and under Turing reductions unless $\textbf{PH}$ collapses at the second level. Ogiwara and Watanabe\cite{OgiwaraWatanabeSparseNP} provide a result employing bounded truth table reductions. Homer and Longpr{\'{e}}\cite{HomerLongpreSparse} provide additional results and alternative proofs. 


\subsection{$\textbf{PH}$ Does Not Collapse}\label{sectionphnoncollapse}
If Conjecture \ref{mainconjecture} holds for stricter versions of $R$ that remain noncomputable even for a TM with an oracle for halting or even for higher levels of $\textbf{AH}$ with iterated oracles for halting, then $\textbf{PH}$ does not collapse, as follows.

Construct these stricter versions of $R$ as follows, with one version hard for each level of $\textbf{AH}$. This definition will use $p$ as both program and input. For language $A$, define the Turing jump of language $A$ as $A'=\{p|p^A(p){\downarrow}\}$, where $A$ is an oracle used in the computation of program $p$ on input $p$. With $\emptyset$ as the empty set, $\emptyset'$ is $\{p|p(p){\downarrow}\}$, and $\emptyset''$ is $p$ such that $p^{\emptyset'}(p)$ halts. Abbreviate $\emptyset^{(0)}{=}\emptyset$ and $\emptyset^{(i+1)}{=}(\emptyset^{(i)})'$. 

Define versions of $R$ hard for each level of $\textbf{AH}$ as follows. Let $R_i{=}\{x|\forall p{:}$ if $|p|{\leq}|x|/2$, then $p^{\emptyset^{(i)}}(p){\nearrow}$ or $p^{\emptyset^{(i)}}(p){\downarrow}$ with $U(p^{\emptyset^{(i)}}(p)){\neq}x\}$. The definition of $R_0$ is the same as $R$ with $p$ replaced by $p(p)$. Proving that $x{\in}R_0$ is typically impossible, so $R_0{\in}\Pi_1^0$ and $R_0{\notin}\Pi_0^0$. Likewise, $R_i{\in}\Pi_{i+1}^0$ and $R_i{\notin}\Pi_i^0$. Although a consistent theory $\mathcal{S}$ with a c.e. set of axioms cannot typically prove $x{\in}R_0$, we can create a theory $\mathcal{S}_1$ that always can by adding $x{\in}R_0$ as a new axiom whenever it is true and $\mathcal{S}\,{\centernot{\sststile{}{}}}`x{\in}R_0$' (equivalently give $\mathcal{S}_1$ a special predicate answering whether $x{\in}R_1$). The set of axioms of $\mathcal{S}_1$ are not c.e. Although  $\mathcal{S}_1\,{{\sststile{}{}}}`x{\in}R_0$' always by construction, typically $\mathcal{S}_1\,{\centernot{\sststile{}{}}}`x{\in}R_1$' if $x{\in}R_1$ and $x{\notin}R_0$. That is, Chaitin's Theorem holds for $\mathcal{S}_1$ and $R_1$. Note that the inclusion $R_{i+1}{\subset}R_i$ is strict, since $\textbf{AH}$ does not collapse. Define $\mathcal{S}_i$ as $\mathcal{S}_{i-1}$ plus new axioms $x{\in}R_i$ whenever that is true and $\mathcal{S}_i\,{\centernot{\sststile{}{}}}`x{\in}R_i$'. By construction, if $x{\in}R_{i-1}$, $\mathcal{S}_i\,{{\sststile{}{}}}`x{\in}R_{i-1}$'.

\begin{theorem}\label{theoremph}
Suppose for each $i{\ge}1$, Conjecture \ref{mainconjecture} holds with $R$ replaced with $R_i$. Then, $\Pi_i^p{\neq}\Pi_{i-1}^p$, and $\textbf{PH}$ does not collapse.
\end{theorem}
\begin{proof}
Let $\mathcal{S}{=}\mathcal{S}_1$. Since Conjecture \ref{mainconjecture} holds with $R$ replaced with $R_1$, there exists $\mathcal{T}$ such that typically, for $x{\in}R_1{\setminus}R_0$, $\mathcal{S}_1\,{\centernot{\sststile{}{t^{\mathcal{O}(1)}}}}``\mathcal{T}\,{\centernot{\sststile{}{t}}}[`x{\in}R_1$'$\,]${''}. Choose one $x{\in}R_1{\setminus}R_0$. However, for any $x$,  $\mathcal{S}_1\,{{\sststile{}{t^{\mathcal{O}(1)}}}}``\mathcal{T}\,{\centernot{\sststile{}{t}}}[`x{\in}R_{0}$'$\,]${''}. The family of sentences $``\mathcal{T}\,{\centernot{\sststile{}{t}}}[`x{\in}R_1$'$\,]${''} indexed by $t$ is in $\Pi_2^p$, while the family of sentences $``\mathcal{T}\,{\centernot{\sststile{}{t}}}[`x{\in}R_{0}$'$\,]${''} is in $\Pi_1^p$. The $x$ chosen above witnesses that $\Pi_2^p{\neq}\Pi_{1}^p$. This argument holds, replacing $R_1$ with $R_i$, for any $i{\geq}1$. Therefore, $\textbf{PH}$ does not collapse.
\end{proof}


\subsection{Folding in Other Open Questions}\label{sectionfoldingin}
The FCI and variants appear to have an unusual capacity to fold in multiple open questions into a single conjecture, that Conjecture \ref{mainconjecture} holds with various definitions of $R$. An interesting question is determine which open questions can and cannot be incorporated in this way. Consider what assumptions would imply exponential-sized circuits for $\texttt{SAT}$ and $\textbf{P=BPP}$. The following argument will require exponential and not just superpolynomial hardness for $\Pi_2^p$ and $\Pi_1^p$. That is, suppose, for $i=1,2$, for $x{\in}R_i$, typically, $\mathcal{S}\,{\centernot{\sststile{}{\Omega(2^{{\delta}t})}}}\mathcal{T}\,{\centernot{\sststile{}{t}}}`x{\in}R_i$' for some $\delta{>}0$. Then, $\texttt{SAT}$ has exponential-sized circuits by a Karp-Lipton argument $\Pi_2^p$, which implies $\textbf{P=BPP}$ (Impagliazzo and Wigderson\cite{Impagliazzo}). Chen et al\cite{ChenRothblumetal} consider versions of the Strong Exponential Time Hypothesis (SETH) sufficient to show $\textbf{P=BPP}$. 

A thorny challenge in complexity theory has been to analyze hardness for small inputs. This is an important practical requirement for instance to be confident that encryption is secure not only asymptotically but for message lengths and key sizes typically used in practice. This analysis is complicated by differences between machine models, and the possibility of adding lookup tables to make any particular small input an easy one. Chaitin's Theorem however does prove impossibility for small inputs, that is, for $|x|{\geq}k$, where $k$ is not very large. It is robust to changes in the details of theories, and holds even if for a theory given lookup tables for instance with axioms asserting $x{\in}R$ for all small $x$ up to any threshold for $|x|$---the Theorem simply yields a larger $k$. 

Consider whether some variants of Conjecture \ref{mainconjecture} would similarly assert hardness for small inputs in complexity theory. The conjecture already asserts hardness for sufficiently long $x$; Conjecture \ref{conjectureisomorphism} will ensure that the same threshold $k$ applies as in Chaitin's Theorem. We would also need to strengthen Conjecture \ref{mainconjecture} so that hardness sets in for small $t$ and not only asymptotically. This latter assumption seems arbitrary. It is also awkward to assert ruling out length $t$ proofs is super-polynomially hard for small $t$, since $t^c>2^t$ for small $t$ with $c$ sufficiently large.  It is more natural to assert a lower bound of $2^{{\delta}t}$ with $\delta{<}1$ to avoid this issue.
\subsection{An Overarching Complexity Conjecture}\label{sectionoverarching}
The following overarching conjecture subsumes all of the conjectures so far in the paper, taking advantage of their similar structure. This yields a single conjecture that multiple open questions have the expected answers.

\begin{conjecture}\label{conjectureoverarching}
For every consistent theory $\mathcal{S}$, there exists a consistent theory $\mathcal{T}$, such that for all $x{\in}R^*$, for some $\delta{<}1$, typically $\mathcal{S}\,{\centernot{\sststile{}{2^{{\delta}t}}}}``\mathcal{T}\,{\centernot{\sststile{}{t}}}[`x{\in}R^*$'$\,]${''}, where $R^*$ is defined with any oracle $O$ for $p$ and assuming compression $|p|{\leq}f(|x|)$ for any decreasing function. 
\end{conjecture}

This conjecture asserts that many open questions have the expected answers:
\begin{theorem}
Conjecture \ref{conjectureoverarching} implies $\textbf{PH}$ does not collapse; $\texttt{TAUT}{\notin}\textbf{AvgP}$; Feige's hypothesis; SETH; exponential-sized circuits for $\texttt{SAT}$;  $\textbf{P=BPP}$; a probabilistic procedure to construct dense hard families of tautologies for a given proof system; and hardness for small inputs.
\end{theorem}

Conjecture \ref{conjectureoverarching} is powerful but messy, but this may be avoided as follows. An intuition for Conjecture \ref{mainconjecture} is that ruling out length $t$ proofs is plausibly hard, given the inaccessibility of the crucial fact that no such proof typically exists of $x{\in}R$. To state this in the most extreme form, suppose no other effectively computable fact about inputs $x{\in}R$ may be useful at all. More formally, there may limited or even nonexistent scope to compute a predicate useful for case-based reasoning about $R$. Similarly, there may be no predicate useful for conditional program logic, making if-then-else statements useless. If this property holds for all but finite $x{\in}R$, then a program can do no better than loops that exhaustively check all cases, and a theory ruling out all length $t$ proofs would need to exhaustively check each case.

\begin{conjecture}\label{conjectureconditional}
Conditional program logic and case-based reasoning are typically useless in ruling out length $t$ proofs that $x{\in}R$. Therfore, the shortest proofs and fastest programs make no or limited use of them and must rely primarily on exhaustive search.
\end{conjecture}

This would be an absurdly strong property for $R$ to have. However, $R$ has other strong properties, for instance, that $R$ passes all known and not yet conceived effective tests of randomness (Li and Vitanyi\cite{LiVitanyiBook} Section 2.4). 

\subsection{Maximally Extending the Isomorphism}\label{sectionextendingisomorphism}
Consider the intriguing possibility that an isomorphism maps not only `$x{\in}R$' but any unprovable sentence to a hard-to-prove sentence. Let $\mathfrak{I}$ be a collection of sentences unprovable in $\mathcal{T}$, initially $\{`x{\in}R$'$|x{\in}R$ and $\mathcal{T}\,{\centernot{\sststile{}{}}}`x{\in}R$'$\}$, and restate Conjecture \ref{mainconjecture} as an assertion about individual unprovable sentences:

\begin{conjecture}\label{conjectureisomorphism}
	For any consistent theories $\mathcal{S}$ and $\mathcal{T}$, with $\mathcal{T}$ strictly stronger than $\mathcal{S}$, if $\phi{\in}\mathfrak{I}$, then: if $\mathcal{S}\,{\centernot{\sststile{}{}}}\mathcal{T}\,{\centernot{\sststile{}{}}}\phi$ then $\mathcal{S}\,{\centernot{\sststile{}{t^{\mathcal{O}(1)}}}}\mathcal{T}\,{\centernot{\sststile{}{t}}}\phi$. 
\end{conjecture}

This conjecture is equivalent to Conjecture \ref{mainconjecture} for $x$ sufficiently large. For small $x$, it asserts exact alignment between unprovable and hard-to-prove sentences related to `$x{\in}R$'. Conjecture \ref{conjectureisomorphism} implies that Conjecture \ref{mainconjecture} holds with ``sufficiently long'' equal to $k$ in Chaitin's Theorem.

Conjecture \ref{conjectureisomorphism} is false if $\mathfrak{I}$ includes all unprovable sentences, by the following counterexample. Choose $x{\in}R$ sufficiently strong that $\mathcal{T}\,{\centernot{\sststile{}{}}}`x{\in}R$', and let $\phi$ be the false unprovable sentence `$x{\notin}R$'. It can be verified in linear time that $\mathcal{T}\,{\centernot{\sststile{}{t}}}`x{\notin}R$', as the finite number of possible witnesses $p$ that `$x{\notin}R$' satisfying $|p|{\leq}|x|/2$ can be easily simulated in parallel to verify $U(p){\neq}x$.\footnote{I appreciate this observation from Pavel Pudlák.}

In that example, `$x{\notin}R$' is false. Suppose this is the reason Conjecture \ref{conjectureisomorphism} fails to hold. It cannot be ruled that Conjecture \ref{conjectureisomorphism} holds for all unprovable sentences that are true. For instance, it could hold for all unprovable sentences `$p{\nearrow}$', or adding alternating quantifiers, for sentences $`\forall y_1\exists y_2\ldots Q y_i U(p,y_1,y_2\ldots y_i){\nearrow}$', where $U$ runs $p$ on inputs $y_i$.

\begin{conjecture}
	Conjecture \ref{conjectureisomorphism} holds for $\mathfrak{I}$ consisting of any true sentences $\phi$ unprovable in $\mathcal{T}$.
\end{conjecture}

This conjecture, if true, would represent an important feature of theories.

\section{Conclusion}\label{conclusion}

The ontological status of Conjecture \ref{mainconjecture} is a high stakes question, as its correctness would resolve multiple open problems. It can be evaluated on the following criteria: falsehood, independence, provability, utility, elegance, and flexibility.

Falsehood is a risk, mitigated by the widespread opportunities to prove such a strong conjecture wrong. An encouraging sign is that Kraj\'{\i}\v{c}ek and Pudlák\cite{Krajicek}'s structurally similar conjecture, $\mathcal{S}\centernot{{{\sststile{}{t^{\mathcal{O}(1)}}}}}$``$\mathcal{T}\,{\centernot{\sststile{}{t}}}[`0{=}1$'\,]{''}, has not been proven false since published in 1989, and an equivalent statement to that, no optimal proof system exists, is widely believed to hold. Furthermore, given the widespread view that most tautologies are hard for most proof systems, it would be surprising if there were not dense witnesses to the nonoptimality of proof systems, as implied by Conjecture \ref{mainconjecture}.

The conjecture's independence from ZFC is also a risk given the extensive reliance on unprovable sentences `$x{\in}R$'. A weaker existential version of Conjecture \ref{mainconjecture} is in fact independent of ZFC: given $\mathcal{S}$, there exists $\mathcal{T}$ and $x{\in}R$ such that $\mathcal{S}{\centernot{\sststile{}{t^{\mathcal{O}(1)}}}}$``$\mathcal{T}{\centernot{\sststile{}{t}}}[`x{\in}R$'$\,]${''}. To see this, create $ZFC^\neg$, an extension of $ZFC$, by adding for any $x$ that is predicted by this weaker conjecture the ``false axiom'' `$x{\notin}R$'.\footnote{$ZFC^\neg$ is not well behaved: It is not arithmetically sound, since `$x{\notin}R$' is false if no $p$ exists as a counterexample to the true sentence `$x{\in}R$'. Also, it has a non-c.e. set of axioms.} This weaker conjecture is false with $\mathcal{S}=ZFC^\neg$, so the conjecture is independent of ZFC.\footnote{This suggests that the existence of optimal proof systems is independent of ZFC, as this is equivalent to the conjecture in footnote \ref{fncoBHP}, which is also existential.} This argument does not work for Conjecture \ref{mainconjecture}, as it would require adding false axioms `$x{\notin}R$' for each $x{\in}R$ w.h.p., contradicting Lemma \ref{lemmahighprobabilityR}. The stronger conjecture may be easier to prove.

Provability is of course an issue for a conjecture implying $\textbf{P}\neq\textbf{NP}$, with many known barriers. If true, the conjecture would be a nonrelativizing fact, requiring a nonrelativizing proof. If the conjecture is true, it limits how oracles operate. Contrary relativizations are not an obstacle in noncomputability theory; rather, adding oracles leads to stronger separations. Conjecture \ref{conjectureisomorphism} forces the same behavior in complexity theory---making $\mathcal{S}$ and $\mathcal{T}$ more powerful strengthens separations on both the noncomputability and complexity sides of the isomorphism. Given that $R$'s pathological behavior includes defeating any conceivable test for randomness, perhaps it also defeats any conceivable conditional program logic or case-based reasoning, per Conjecture \ref{conjectureconditional}.

Utility is also a consideration---the conjectures may be useful as working hypotheses. They allow unrelated open questions to be viewed in a common framework. They may also analysis of a question to swap back and forth between noncomputability theory and complexity theory.

Elegance is also an attraction. If a theory's impossible-to-prove and hard-to-prove sentences are isomorphic, this would reveal a hidden symmetry, potentially with implications for mathematics outside complexity theory.

The conjecture's flexibility is notable. A clearer sense is needed regarding when the conjecture might or might not hold to avoid arbitrary choices. 

The paper points to numerous opportunities for further research. First, this could to relate FCI to other open questions and hardness hypotheses. Second, it would be helpful to clarify what is the underlying principle driving Conjecture \ref{conjectureoverarching}, that that $R$ frustrates conditional program logic and case-based reasoning, per Conjecture \ref{conjectureconditional}, implying FCI. Finally, the conjecture that an isomorphism maps any true unprovable sentence to a hard-to-prove family of sentences is appealing as a new potential structure feature of theories such as ZFC.

\bibliographystyle{amsplain}
\bibliography{equivalence}

\end{document}